\documentclass[12pt]{article}

\title{On the mystery of quantum probabilistic rule: trigonometric and hyperbolic
probabilistic behaviours}
\author{Andrei Khrennikov\\
Department of Mathematics, Statistics and Computer Sciences\\
University of V\"axj\"o, S-35195, Sweden}
\begin{document}
\maketitle

\begin{abstract} We demonstrate that the origin of 
so called quantum probabilistic rule (which differs from
the classical Bayes' formula by the presence of $\cos \theta$-factor) might be explained
in the framework of ensemble fluctuations which are induced 
by preparation procedures. In particular, quantum rule for probabilities (with nontrivial
$\cos \theta$-factor) could be simulated for macroscopic physical systems via preparation
procedures producing ensemble fluctuations
of a special form. We discuss preparation and measurement 
procedures which may produce probabilistic rules which are neither classical nor quantum;
in particular, hyperbolic `quantum theory.'
\end{abstract}

\section{Introduction}

It is well known that the classical probabilistic rule based on the Bayes' 
formula for conditional probabilities cannot be applied to quantum formalism, 
see, for example, [1]-[3] for extended discussions. In fact, all special properties of quantum systems 
are just consequences of violations of the classical probability rule, Bayes' theorem [1]. In this paper
we restrict our investigations to the two dimensional case. Here Bayes' formula has the form $(i=1,2):$
\begin{equation}
\label{*1}
{\bf p}(A=a_i)={\bf p}(C=c_1){\bf p}(A=a_i/C=c_1)+
{\bf p}(C=c_2){\bf p}(A=a_i/C=c_2), 
\end{equation}
where $A$ and $C$ are physical variables which take, respectively, values  $a_1, a_2$ and $c_1, c_2.$
Symbols ${\bf p}(A=a_i/C=c_j)$ denote conditional probabilities.
There is  a large diversity of opinions on the origin of violations of (\ref{*1}) in quantum mechanics. 
The common opinion is that violations of (\ref{*1}) are induced by special properties of quantum systems. 

Let $\phi$ be a quantum state. Let $\{\phi_i\}_{i=1}^2$ be an orthogonal basis consisting of 
eigenvectors of the operator $\hat{C}$ corresponding to the physical observable $C.$
The quantum theoretical rule has the form $(i=1,2):$
\begin{equation}
\label{*2}
q_i = {\bf p}_1 {\bf p}_{1i} + {\bf p}_2 {\bf p}_{2i}\pm 
2 \sqrt{{\bf p}_1{\bf p}_{1i} {\bf p}_2 {\bf p}_{2i}}\cos\theta,
\end{equation}
where
$q_i={\bf p}_\phi(A=a_i), {\bf p}_j={\bf p}_\phi(C=c_j), {\bf p}_{ij}={\bf p}_{\phi_i}(A=a_j), i,j=1,2.$
Here probabilities have indexes corresponding to quantum states.
The common opinion is that this quantum probabilistic rule must be considered as a peculiarity of nature.
However, there exists an opposition to this general opinion, namely the probabilistic opposition. 
The main domain of activity of this probabilistic opposition is Bell's inequality 
and the EPR paradox [4] , see, for example, [1], [5]-[11]. 
The general idea supported by the probabilistic opposition is that special 
quantum behaviour can be understood on  the basis of 
local realism, if we be careful with the probabilistic description of physical phenomena. 
It seems that the origin of all `quantum troubles' is probabilistic rule (\ref{*2}). 
It seems that the violation of Bell's inequality is just a new representation of the old contradiction
between rules (\ref{*1}) and (\ref{*2}) (the papers of Accardi [1] and  
De Muynck, De Baere and Martens [7] contain extended discussions on this problem).
Therefore, the main problem of the probabilistic justification of quantum mechanics is to find the clear probabilistic 
explanation of the origin of quantum probabilistic rule (\ref{*2}) and the violation of 
classical probabilistic rule (\ref{*1}) and explain why (\ref{*2}) is sometimes  reduced to (\ref{*1}).

L. Accardi [5] introduced a notion of the {\it statistical invariant} to investigate the relation between
classical Kolmogorovean and quantum probabilistic models, see also Gudder and Zanghi in [6]. 
He was also the first who 
mentioned that Bayes' postulate is a "hidden axiom of the Kolmogorovean model... which limits its applicability to 
the statistical description of the natural phenomena ", [5]. In fact, this investigation plays a crucial
role in our analysis of classical and quantum probabilistic rules.

An interesting investigation on this problem is contained in the paper of J. Shummhammer [11].
He supports the idea that quantum probabilistic rule (\ref{*2}) is not a peculiarity of 
nature, but just a consequence of one special method of the probabilistic description of nature,
so called method of {\it {maximum predictive power}}. We do not directly support the idea of Shummhammer. 
It seems that the origin of (\ref{*2}) is not only a consequence of the use of one 
special method for the description of nature, but merely a consequence of our manipulations with nature, 
ensembles of physical systems, in quantum preparation/measurement procedures. 

In this paper we provide probabilistic analysis of quantum rule (\ref{*2}). In our analysis `probability'
has the meaning of the {\it frequency probability,} namely the limit of frequencies in a long 
sequence of trials (or for a large statistical ensemble). Hence, in fact , we follow to 
R. von Mises' approach to probability [12]. It seems that it would be 
impossible to find the roots of quantum rule (\ref{*2}) in the conventional probability framework, 
A. N. Kolmorogov, 1933, [13]. In the conventional measure-theoretical framework probabilities are defined as sets of real numbers having some special mathematical properties.
Classical rule (\ref{*1}) is merely a consequence of the definition of conditional probabilities. 
In the Kolmogorov framework
to analyse the transition from (\ref{*1}) to (\ref{*2}) is to analyse the transition from one definition to another. 
In the frequency framework we can analyse behaviour of trails which induce one or another property
of probability. Our analysis shows that quantum probabilistic rule (\ref{*2}) can be explained on 
the basis of ensemble fluctuations (one of possible sourses of ensemble fluctuations is
so called ensemble nonreproducibility, see De Baere [7];
see also [10] for the statistical variant of nonreproducibility). Such fluctuations can generate (under special conditions) 
the cos $\theta$-factor in (\ref{*2}). Thus trigonometric fluctuations of quantum probabilities can 
be explained without using the wave arguments.

An unexpected consequence of our analysis is that quantum probability rule (\ref{*2}) is just one of 
possible perturbations (by ensemble fluctuations) of classical probability rule (\ref{*1}). In principle,
there might exist experiments which would produce perturbations of classical probabilistic rule (\ref{*1})
which differ from quantum probabilistic rule (\ref{*2}).

\section{Quantum formalism and ensemble fluctuations}

{\bf 1. Frequency probability theory.} The frequency definition of probability is more or less standard in quantum theory;
especially in the approach based on preparation and measurement procedures, [14], [3]. 

 Let us consider a sequence of physical systems
$
\pi= (\pi_1,\pi_2,..., \pi_N,... )\;.
$
Suppose that elements of $\pi$ have some property, for example, position, 
and this property
can be described by  natural numbers: $ L=\{1,2,...,m \},$ 
the set of labels.  Thus, for each $\pi_j\in \pi,$ we have a number $x_j\in L.$
So $\pi$ induces a sequence 
\begin{equation}
\label{la1}
x=(x_1,x_2,..., x_N,...), \; \; x_j \in L.
\end{equation}
For each fixed $\alpha \in L,$ we have the relative frequency
$\nu_N(\alpha)= n_N(\alpha)/N$ of the appearance of $\alpha$
in $(x_1,x_2,..., x_N).$ Here $n_N(\alpha)$ is the number of elements in
$(x_1,x_2,..., x_N)$ with $x_j=\alpha.$
R. von Mises [12] said that
$x$ satisfies to the principle of the {\it statistical stabilization} of relative frequencies,
if, for each fixed $\alpha \in L,$ there exists the
limit 
\begin{equation}
\label{l0}
{\bf p} (\alpha)=\lim_{N\to \infty} \nu_N(\alpha) .
\end{equation}
This limit is said to be a probability of $\alpha.$

We shall not consider so called principle of {\it randomness,} see
[12] for the details. This principle, despite its importance for the foundations of probability theory, is 
not related to
our frequency analysis. We shall be interested only in the statistical stabilization of relative frequencies.

{\bf 2. Preparation and measurement procedures and quantum formalism.}
We consider a statistical ensemble S of quantum particles described by a quantum state $\phi.$ 
This ensemble is produced by some preparation procedure ${\cal E},$ see, for example, [14], [3] for details.
There are two discrete physical observables $C=c_1, c_2$ and $A=a_1, a_2.$

The total number of particles in S is equal to N.
Suppose that $n_{i}^{c}, i=1,2,$ particles in $S$ would give the result 
$C=c_i$ and $n_{i}^a, i=1,2,$ particles in $S$ would give the result 
$A=a_i.$ Suppose that, among those particles which would produce $C=c_i,$ 
there are $n_{ij}, i,j, =1,2,$ particles which would give the result $A=a_j$
(see (R) and (C) below to specify the meaning of `would give'). So 

$n_i^c=n_{i1}+n_{i2}, n_j^a=n_{1j}+n_{2j}, i, j=1,2.$

\medskip

(R) We can use an objective realist 
model in that both $C$ and $A$ are {\it objective properties}
of a quantum particle, see [2], [3], [10] for the details. 
In such a model we can consider in $S$ sub-ensembles ${\rm{S_j(C)}}$ and ${\rm{S_j(A), j=1, 2, }}$ 
of particles having properties ${\rm{C=c_j}}$ and $A=a_j,$ respectively. Set 
${\rm{S_{ij}(A,C)=S_i(C)\cap S_j(A).}}$ Then $n_{ij}$ is 
the number of elements in the ensemble ${\rm{S_{ij}(A,C).}}$ We remark 
that the `existence' of the objective property $(C=c_i$ and $A=a_j)$ need not imply
the possibility to measure this property. For example, such a measurement is impossible
in the case of incompatible observables. So in general $(C=c_i$ and $A=a_j)$ is a kind of hidden property.

\medskip

(C) We can use so called {\it contextualist} realism, see, for example, [3] and
De Muynck, De Baere and Martens in [7]. 
Here we cannot assume that a quantum system determines uniquely the result of a measurement. 
This result depends not only on properties of a quantum particle, but also  on 
the experimental arrangement. Here $n_{ij}$ is the number of particles 
which would produce $C=c_i$ and $A=a_j.$ We remark that the latter statement, in fact, contains
{\it counterfactuals}: $C=c_i$ and $A=a_j$ could not be measured simultaneously, see, for example, [3] for 
the use of counterfactuals in quantum theory.

\medskip

The quantum experience says that the following frequency probabilities are well defined for
all observables $C,A :$
\begin{equation}
\label{*3C}
{\bf p}_i={\bf p}_\phi(C=c_i)=\lim_{N\rightarrow\infty}{\bf p}_i^{(N)}, {\bf p}_i^{(N)}=
{\frac{n_i^c}{N}}; 
\end{equation}
\begin{equation}
\label{*3A}
q_i={\bf p}_\phi(A=a_i)=\lim_{N\rightarrow\infty} q_i^{(N)}, q_i^{(N)}={\frac{n_i^a}{N}}.
\end{equation}
Can we say something about behaviour of frequencies 
$\rm{\tilde{{\bf p}}_{ij}^{(N)}={\frac{n_{ij}}{N}}, N\rightarrow\infty}$?
Suppose that they stabilize, when ${\rm{N\rightarrow \infty.}}$ This implies that probabilities
$\tilde{{\bf p}}_{{\rm{ij}}}={\bf p}_\phi({\rm{C=c_i,A=a_j}})=\lim_{N\rightarrow\infty}
\tilde{{\bf p}}_{{\rm{ij}}}^{({\rm{N}})}$
would be well defined. The quantum experience says that (in general) this is not the case. Thus, in
general, the frequencies $\tilde{{\bf p}}_{ij}^{(N)}$ fluctuate, when ${\rm{N\rightarrow\infty.}}$
Such fluctuations can, nevertheless, produce the statistical stabilization (\ref{*3C}), (\ref{*3A}), 
see [10] for the details.

{\bf Remark 2.1.} {\small The common interpretation of experimental violations of Bell's inequality 
is that realism and even contextualist realism cannot be used in quantum theory 
(at least in the local framework). However, Bell's considerations only imply 
that we cannot use realist models under the assumption that $\rm{\tilde{{\bf p}}_{ij}^{(N)}}$ stabilize.
The realist models with fluctuating frequencies $\rm{\tilde{{\bf p}}_{ij}^{(N)}}$ can coexist with violations 
of Bell's inequality, see [10].}

Let us now consider statistical ensembles $T_i, i=1,2,$ of quantum particles 
described by the quantum states $\phi_{i}$ which are eigenstates of the operator 
$\hat{C}:$ $\hat{C} \phi_i =c_i \phi_i.$
These ensembles are produced by some preparation procedures ${\cal E}_i.$ 
For instance, we can suppose that particles produced by a preparation procedure ${\cal E}$
for the quantum state $\phi$ pass through additional 
filters $F_i, i=1,2.$ In quantum formalism we have 
\begin{equation}
\label{*4}
\phi=\sqrt{{\bf p}_1} \; \phi_1+\sqrt{{\bf p}_2} e^{i \theta}  \; \phi_2 \;.
\end{equation}
In the objective realist model (R) this representation may induce the illusion that ensembles $T_i, i=1,2,$ 
for states $\phi_i$ must be identified with 
sub-ensembles ${\rm{S_i(C)}}$ of the ensemble $S$ for the state $\phi.$ 
However, there are no physical reasons for such an identification.
There are two main sources of troubles with this identification:

\medskip

(a). The additional filter ${\rm{F}}_1$ (and ${\rm{F}}_2)$ changes the properties of quantum particles. 
The probability distribution of the property $A$ for the
ensemble ${\rm{S_1(C)=\{{\pi\in S:C(\pi)=c_1}}}\}$ 
(and ${\rm{S_2(C))}}$ may differ from the corresponding probability distribution for the ensemble ${\rm{T_1}} ($and ${\rm{T_2).}} $ So different preparation procedures produce different distributions of properties. The same conclusion can be done for the contextualist realism: an additional filter changes possible reactions of quantum particles to measurement devices. 

\medskip

(b). As we have already mentioned, frequencies ${\rm{\tilde{{\bf p}}_{ij}^{(N)}={\frac{n_{ij}}{N}}}}$ must fluctuate 
(in the case of incompatible observables). 
Therefore, even if additional filters do not change properties of quantum particles,
nonreproducibility implies that the distribution of the property $A$ 
may be essentially different for statistical ensembles ${\rm{S_1(C) }}$ and ${\rm{S_2(C) }}$
(sub-ensembles of $S)$ and ${\rm{T}}_1$ and ${\rm{T_2}}.$
Moreover, distributions may be different even for sub-ensembles ${\rm{S_1(C)}}$ and ${\rm{S}_1^\prime(C)}$ 
(or ${\rm{S_2(C)}}$ and ${\rm{S_2^\prime(C)),}} $ of two different ensembles S and ${\rm{S^\prime}}$ of
quantum particles prepared in the same quantum state $\phi,$ see [10].

\medskip

Fluctuations of physical properties which could be induced by (a) or (b)
will be called {\it{ensemble fluctuations.}}

Suppose that ${\rm{m_{ij}}}$ particles in the ensemble ${\rm{T_i}}$ would produce the result 
$A=a_j, j=1,2.$ We can use the objective realist model, (R). Then $m_{ij}$ is just the number of particles in the 
ensemble $T_i$ having the objective property $A=a_j.$ We can also use the contextualist 
model,  (C). Then $m_{ij}$ is the number of particles  in the ensemble $T_i$ which in the
process of an interaction with 
a measurement device for the physical observable $A$ would give the result $A=a_j.$

The quantum experience says that the following frequency probabilities are well defined:

${\rm{{\bf p}_{ij}={\bf p}_{\phi_i}(A=a_j)=lim_{N\rightarrow\infty}
{\bf p}_{ij}^{(N)}, {\bf p}_{ij}^{(N)}={\frac{m_{ij}}{n_i^c}}.}}$ 

Here it is assumed that an ensemble ${\rm{T_i}}$ consists of ${\rm{n_i^c}}$ particles, $i=1,2.$ 
It is also assumed that ${\rm{n_i^c=
n_i^c (N) \rightarrow \infty, N\rightarrow\infty.}}$ In fact, the latter assumption holds true 
if both probabilities ${\bf p}_i, i=1,2,$ are nonzero.

We remark that probabilities ${\bf p}_{ij}={\bf p}_{\phi_i}(A=a_j)$ cannot be (in general)
identified with conditional probabilities ${\bf p}_\phi(A=a_j/C=c_i)={\frac{\tilde{{\bf p}}_{ij}}{{\bf p}_i}}.$ 
As we have remarked, these probabilities are related to statistical ensembles prepared by different 
preparation procedures, namely by ${\cal E}_i, i=1,2,$ and ${\cal E}.$

Let $\{ \psi_j \}_{j=1}^2$ be an orthonormal basis consisting of 
eigenvectors of the operator $A.$ We can restrict our considerations to the case:
\begin{equation}
\label{*4a}
\phi_1 = \sqrt{{\bf p}_{11}} \;\psi_1 + e^{i\gamma_1} \sqrt{{\bf p}_{12}}\; \psi_2, \;\;
\phi_2= \sqrt{{\bf p}_{21}} \;\psi_1 + e^{i\gamma_2} \sqrt{{\bf p}_{22}} \;\psi_2\;.
\end{equation}
As $(\phi_1, \phi_2)=0,$ we obtain:

$\rm{{\sqrt{{\bf p}_{11}{\bf p}_{21}}} + e^{i(\gamma_1 - \gamma_2)} {\sqrt{{\bf p}_{12} {\bf p}_{22}}} =0}.$ 

Hence, $\sin(\gamma_1-\gamma_2)=0$ (we
suppose that all probabilities ${\bf p}_{ij} >0)$ 
and $\gamma_2=\gamma_1+\pi k. $ 
We also have

${\sqrt{{\bf p}_{11}{\bf p}_{21}}}
+\cos(\gamma_1-\gamma_2){\sqrt{{\bf p}_{12}{\bf p}_{22}}}=0.$

This implies that $k=2 l+1$ and ${\sqrt{{\bf p}_{11} {\bf p}_{21}}}={\sqrt{{\bf p}_{12} {\bf p}_{22}}}.$ 
As ${\bf p}_{12}=1-{\bf p}_{11} $ and $ {\bf p}_{21}=1-{\bf p}_{22},$ 
we obtain that 
\begin{equation}
\label{*4b}
{\bf p}_{11}={\bf p}_{22}, \; \; {\bf p}_{12}={\bf p}_{21}.
\end{equation}
This equalities are equivalent to the condition: ${\bf p}_{11}+{\bf p}_{21}=1, 
{\bf p}_{12}+{\bf p}_{22}=1.$
So the matrix of probabilities $({\bf p}_{ij})_{i,j=1}^2$ is so called {\it double
stochastic matrix,} see, for example, [3] for general considerations.

Thus, in fact, 
\begin{equation}
\label{*q}
\phi_1={\sqrt{{\bf p}_{11}}}\;\psi_1+e^{i\gamma_1}{\sqrt{{\bf p}_{12}}}\;\psi_2,\;
\phi_2={\sqrt{{\bf p}_{21}}}\;\psi_1-e^{i\gamma_1}{\sqrt{{\bf p}_{22}}}\;\psi_2.
\end{equation}
So $\rm{\varphi=d_1\psi_1+d_2\psi_2,}$ where

$d_1=\sqrt{{\bf p}_{1} {\bf p}_{11}} + e^{i\theta} \sqrt{{\bf p}_{2} {\bf p}_{21}}, \;
d_2=e^{i\gamma_1} \sqrt{{\bf p}_{1} {\bf p}_{12}} - e^{i(\gamma_1+\theta)}
\sqrt{{\bf p}_{2} {\bf p}_{22}}.$ 

Thus 
\begin{equation}
\label{*}
q_1={\bf p}_\phi (A=a_1)=|d_1|^2={\bf p}_1{\bf p}_{11}+{\bf p}_2{\bf p}_{21}+ 2 \sqrt{{\bf p}_1 {\bf p}_{11}
{\bf p}_2 {\bf p}_{21}}\cos \theta ;
\end{equation}
\begin{equation}
\label{**}
q_2= {\bf p}_\phi(A=a_2) = |d_2|^2= {\bf p}_1 {\bf p}_{12}
+{\bf p}_2 {\bf p}_{22} -2 \sqrt{{\bf p}_1 {\bf p}_{12} {\bf p}_2 {\bf p}_{22}} \cos\theta .
\end{equation}

{\bf 3. Probablity relations connecting preparation procedures.}
Let us forget at the moment about the quantum theory. We consider an arbitrary 
preparation procedure ${\cal E}$ for microsystems or macrosystems. Suppose that
${\cal E}$ produced an ensemble $S$ of physical systems. Let $C (=c_1, c_2)$ 
and $A (= a_1, a_2)$ be physical quantities which can be measured for elements 
$\pi \in S.$ Let ${\cal E}_1$ and ${\cal E}_2$ be preparation procedures which are based on filters
$F_1$ and $F_2$ corresponding, respectively, to values $c_1$ and $c_2$ of $C.$ Denote 
statistical ensembles produced by these preparation procedures by symbols $T_1$ and $T_2,$
respectively. Symbols $N, n_i^c, n_i^a, n_{ij}, m_{ij}$ have the same meaning as in the previous
considerations. Probablities  ${\bf p}_i, {\bf p}_{ij}, q_i$ are defined in the same way as
in the previous considerations. The only difference is that, instead of indexes corresponding to quantum states, we use indexes corresponding to 
statistical ensembles: ${\bf p}_i= {\bf P}_S (C=c_i), q_i= {\bf P}_S (A=a_i), {\bf p}_{ij}= {\bf P}_{T_i} (A=a_i).$

In the classical frequency framework we obtain:
$$
q_1^{(N)} = \frac{n_1^a}{N} = \frac{n_{11}}{N} + \frac{n_{21}}{N} =
\frac{m_{11}}{N} + \frac{m_{21}}{N} + \frac{(n_{11} - m_{11})}{N} + \frac{(n_{21} - m_{21})}{N}.
$$

But, for  $i=1,2,$ we have

$\frac{m_{1i}}{N} = \frac{m_{1i}}{n_{1}^c}\cdot \frac{n_{1}^c}{N}
={\bf p}_{1i}^{(N)} {\bf p}_1^{(N)},\; \; 
\frac{m_{2i}}{N} =\frac{m_{2i}}{n_2^c}\cdot \frac{n_{2}^c}{N}= {\bf p}_{2i}^{(N)}{\bf p}_2^{(N)}\;.
$

Hence
\begin{equation}
\label{a*}
q_i^{(N)} = {\bf p}_{1}^{(N)}  {\bf p}_{1i}^{(N)} + {\bf p}_{2}^{(N)} {\bf p}_{2i}^{(N)} + \delta_i^{(N)},
\end{equation}
where
$$
\delta_i^{(N)} = \frac{1}{N} [(n_{1i} - m_{1i}) + (n_{2i} - m_{2i})],  \; i= 1,2.
$$
In fact, this rest term depends on the statistical ensembles
$S, T_1, T_2,$ $\delta_i^{(N)} = \delta_i^{(N)}(S,T_1,T_2).$ 

{\bf 4. Behaviour of fluctuations.}
First we remark that $\lim_{N\rightarrow\infty} \delta_i^{(N)}$ exists for all physical 
measurements. This is a consequence of the property of statistical stabilization of 
relative frequencies for physical observables (in classical as well as in 
quantum physics).
It may be that this property is a peculiarity of nature. 
It may be that this is just a property of our measurement and preparation procedures, see [10] 
for an extended discussion. In any case we always observe that

${\rm{q_i^{(N)}\rightarrow q_i, {\bf p}_i^{(N)}\rightarrow {\bf p}_i, {\bf p}_{ij}^{(N)}\rightarrow {\bf p}_{ij}, N\rightarrow\infty.}}$ 

Thus there exist limits

$\delta_i =\lim_{N\rightarrow\infty}\delta_i^{(N)}=q_i-{\bf p}_1{\bf p}_{1i}-{\bf p}_2{\bf p}_{2i}.$ 

Suppose
that ensemble fluctuations produce negligibly small (with respect to N) changes in properties of
particles. Then 
\begin{equation}
\label{cl}
\delta_i^{(N)}\rightarrow 0, N\rightarrow \infty.
\end{equation}
This asymptotic implies  classical probablistic rule (\ref{*1}). 
In particular, this rule appears in all experiments of classical physics. Hence, preparation and measurement
procedures of classical physics produce ensemble fluctuations with asymptotic (\ref{cl}).
We also have  such a behaviour
in the case of compatible observables  in quantum physics.
Moreover, the same classical probabilistic rule we
can obtain for incompatible observables $C$ and $A$ if the phase factor $\theta= \frac{\pi}{2} + \pi k.$
Therefore classical probabilistic rule (\ref{*1}) is not directly related to commutativity of corresponding 
operators in quantum theory. It is a consequence of asymptotic (\ref{cl}) for ensemble fluctuations.

Suppose now that
filters $F_i, i =1,2,$ produce relatively large (with respect to N) changes in properties of
particles. Then
\begin{equation}
\label{*c}
 \lim_{N\rightarrow\infty} \delta_i^{(N)}= \delta_i \not= 0. 
\end{equation}
Here we obtain probabilistic rules which differ from the classical one, (\ref{*1}).
In particular, this implies that behaviour of ensemble fluctuations (\ref{*c}) cannot 
be produced in experiments of classical physics. A rather special class of ensemble fluctuations (\ref{*c})
is produced in experiments of quantum physics. However, ensemble fluctuations of form (\ref{*c})
are not reduced to quantum fluctuations (see further considerations).

To study carefully behaviour of fluctuations $\delta_i^{(N)},$ we represent them as:

$\delta_i^{(N)}=2 \sqrt{{\bf p}_1^{(N)} {\bf p}_{1i}^{(N)} {\bf p}_2^{(N)} {\bf p}_{2i}^{(N)}}
\lambda_i^{(N)},
$

where 

$\lambda_i^{(N)} = \frac{1}{2 \sqrt{m_{1i} m_{2i}}}
[(n_{1i} - m_{1i}) + (n_{2i} - m_{2i})]\;.
$

We have used the fact: 

${\bf p}_1^{(N)} {\bf p}_{1i}^{(N)} {\bf p}_2^{(N)} {\bf p}_{2i}^{(N)}
=
\frac{n_1^c}{N} \cdot \frac{m_{1i}}{n_1^c} \cdot\frac{n_2^c}{N} \cdot
\frac{m_{2i}}{n_{2}^c}
=\frac{m_{1i} m_{2i}}
{N^2}.
$ 

We have:
$\delta_i
=2  \sqrt{{\bf p}_1 {\bf p}_{1i} {\bf p}_2 {\bf p}_{2i}} \;\lambda_{i},$ where the coefficients
$\lambda_i=\lim_{N\rightarrow\infty} \lambda_i^{(N)}, i=1,2.$

In classical physics the coefficients $\rm{\lambda_i=0.}$ The same situation 
we have in quantum physics for all compatible
observables as well as for some incompatible observables.
In the general case in quantum physics we can only say that
\begin{equation}
\label{*q1}
|\lambda_i|\leq 1 .
\end{equation}

Hence, for quantum fluctuations, we always have:
$$
|{\frac{(n_{1i}-m_{1i})+(n_{2i}-m_{2i})}{2{\sqrt{m_{1i}m_{2i}}}}}|\leq 1, N \rightarrow \infty.
$$
Thus quantum ensemble fluctuations induce a relatively small (but in general nonzero!) variations
of properties. 

{\bf 4. Fluctuations which induce the quantum probabilistic rule.}
Let us
consider preparation procedures ${\cal E}, {\cal E}_j, j=1,2,$ which have the deviations, 
when $N\rightarrow\infty$, of the following form $(i=1,2):$
\begin{equation}
\label{*7}
\epsilon_{1i}^{(N)}= n_{1i}-m_{1i}= 2 \xi_{1i}^{(N)} \sqrt{ m_{1i} m_{2i}}, 
\end{equation}
\begin{equation}
\label{*8}
\epsilon_{2i}^{(N)}= n_{2i}-m_{2j} = 2 \xi_{2i}^{(N)} \sqrt{m_{1i}m_{2i}}, 
\end{equation}

where the coefficients $\xi_{ij}$ satisfy the inequality
\begin{equation}
\label{*8a}
|\xi_{1i}^{(N)}+\xi_{2i}^{(N)}|\leq1, N\rightarrow\infty.
\end{equation}

Suppose that ${\rm{\lambda_i^{(N)}=\xi_{1i}^{(N)}+\xi_{2i}^{(N)}\rightarrow\lambda_i, N\rightarrow\infty,}}$
where ${\rm{|\lambda_i|\leq1.}}$ We can represent ${\rm{\lambda_i^{(N)}=\cos \theta_i^{(N)}.}}$ 
Then $\theta_i^{(N)}\rightarrow\theta_i, \rm{mod} 2 \pi,$ when  $N\rightarrow\infty.$ Thus ${\rm{\lambda_i=\cos\theta_i.}}$ 

We obtained that:
\begin{equation}
\label{*N}
\delta_i=2 \sqrt{ {\bf p}_1 {\bf p}_{1i} {\bf p}_2 {\bf p}_{2i}}
\cos\theta_i, i=1,2.
\end{equation}
Thus fluctuations of the form (\ref{*7}), (\ref{*8}) produce 
the probability rule $(i=1,2):$
\begin{equation}
\label{*n}
q_i={\bf p}_1{\bf p}_{1i}+{\bf p}_2{\bf p}_{2i}+2{\sqrt{{\bf p}_1{\bf p}_2{\bf p}_{1i}{\bf p}_{2i}}}\cos\theta_i.
\end{equation}

The usual probabilistic calculations  give us

$
1 = q_1 + q_2 = {\bf p}_1 {\bf p}_{11} + 
{\bf p}_2{\bf p}_{21} + +{\bf p}_1{\bf p}_{12}+{\bf p}_2{\bf p}_{22}+
$

$
2\; \sqrt{{\bf p}_1 {\bf p}_2 {\bf p}_{11} {\bf p}_{21}}
\cos\theta_1+
2\; \sqrt{{\bf p}_1{\bf p}_2{\bf p}_{12}{\bf p}_{22}} \cos \theta_2
$

$
=1+2 \sqrt{{\bf p}_1 {\bf p}_2}
[\sqrt{{\bf p}_{11}{\bf p}_{21}} \cos\theta_1+ \sqrt{{\bf p}_{12}{\bf p}_{22}}\cos\theta_2]\;.
$

Thus we obtain the relation:
\begin{equation}
\label{*N2}
\sqrt{{\bf p}_{11} {\bf p}_{21}} \cos\theta_1 + \sqrt{{\bf p}_{12} {\bf p}_{22}}  \cos\theta_2 =0 \;. 
\end{equation}
Suppose that ensemble fluctuations (\ref{*7}), (\ref{*8}), satisfy the additional condition
\begin{equation}
\label{*N1}
\lim_{N\rightarrow\infty}{\bf p}_{11}^{(N)}=
\lim_{N\rightarrow\infty}{\bf p}_{22}^{(N)}.
\end{equation}
This condition implies that the matrix of probabilities is a double stochastic matrix.
Hence, we get
\begin{equation}
\label{*N3}
\cos\theta_1= -\cos\theta_2\;.
\end{equation}
So we demonstrated that ensemble fluctuations (\ref{*7}), (\ref{*8}) in the combination with
double stochastic condition (\ref{*N1}) produce quantum probabilistic relations (\ref{*}), (\ref{**}).

It must be noticed that the existence of the limits
$\lambda_i=\lim_{N\rightarrow\infty}\lambda_i^{(N)}$ 
does not imply the existence of limits
$\xi_{1i}=\lim_{N\rightarrow\infty}\xi_{1i}^{(N)}$ and $\xi_{2i}=\lim_{N\rightarrow\infty}\xi_{2i}
^{(N)}.$ For example, let $\xi_{1i}^{(N)}
=\lambda_i \cos^2\alpha_i^{(N)}$ and $\xi_{2i}^{(N)}=
\lambda_i\sin^2\alpha_i^{(N)}$, 
where `phases' $\alpha_i^{(N)}$ fluctuate $\rm{mod}\;\; 2\pi.$ Then
numbers $\rm{\xi_{1i}}$ and $\rm{\xi_{2i}}$ are not defined, but $\rm{\lim_{N\rightarrow\infty} [\xi_{1i}^{(N)}
+\xi_{2i}^{(N)}]=\lambda_i, i=1,2,}$ exist.

If $\xi_{ij}^{(N)}$ stabilize, then probabilities for the simultaneous measurement of incompatible
observables would be well defined: 

${\bf p} (A=a_1, C=c_1)
=\lim_{N\rightarrow\infty}
\frac{n_{11}}{N}=
{\bf p}_1 {\bf p}_{11} + 2 \sqrt{{\bf p}_1 {\bf p}_2 {\bf p}_{11} {\bf p}_{21}}
\xi_{11}, \ldots.
$ 

The quantum formalism implies that in general such probabilities do not exist.

{\bf Remark 2.1.} The magnitude of fluctuations can be found experimentally. Let C and A be two physical
observables. We prepare free statistical ensembles $\rm{S, T_1, T_2}$ 
corresponding to states $\phi,
\phi_1, \phi_2.$ By measurements of C and A for $\rm{\pi\in S}$ we obtain frequencies $\rm{{\bf p}_1^{(N)},
{\bf p}_2^{(N)}, q_1^{(N)}, q_2^{(N)},}$ by measurements of A for $\rm{\pi\in T_1}$ and for $\rm{\pi\in T_2}$ we
obtain frequencies $\rm{{\bf p}_{1i}^{(N)}.}$ We have 

$\rm{f_i(N)=\lambda_i^{(N)}={\frac{q_i^{(N)}-{\bf p}_1^{(N)}
{\bf p}_{1i}^{(N)}-{\bf p}_2^{(N)}{\bf p}_{2i}^{(N)}}{2\;\sqrt{{\bf p}_1^{(N)}{\bf p}_{1i}^{(N)}{\bf p}_2^{(N)}
{\bf p}_{2i}^{(N)}}}}}$

It would be interesting to obtain graphs of functions $\rm{f_i(N)}$ for different pairs of 
physical observables. Of course, we know that $\lim_{N\to \infty} f_i(N)= \pm \cos\theta.$ However,
it may be that such graphs can present a finer structure of quantum states.

\section{On the magnitude of fluctuations which produce the classical probabilistic rule}

We remark that the classical probabilistic rule (which is induced by
ensemble fluctuations with $\xi_i^{(N)}\rightarrow 0)$ can be observed for
fluctuations having relatively large absolute magnitudes. For instance, let 
\begin{equation}
\label{*9} 
\epsilon_{1i}^{(N)}=2\xi_{1i}^{(N)}
\sqrt{m_{1i}},\; \;  \epsilon_{2i}^{(N)}=2\xi_{2i}^{(N)}
\sqrt{m_{2i}}, i=1,2,
\end{equation}
where sequences of coefficients $\{\xi_{1i}^{(N)}\}$ 
and 
$\{\xi_{2i}^{(N)}\}$ are bounded 
$(N\rightarrow\infty).$ 
Here 

$\lambda_i^{(N)}=
\frac{\xi_{1i}^{(N)}}
{\sqrt{m_{2i}}} + \frac{\xi_{2i}^{(N)}}{m_{1i}}
\rightarrow 0, N\rightarrow\infty
$ 

(as usual, we assume that ${\bf p}_{ij} >0).$

{\bf Example 3.1.} Let $N\approx 10^6, n_1^c \approx n_2^c\approx 5\cdot 10^5, 
m_{11}\approx m_{12}\approx m_{21}\approx m_{22}\approx 25\cdot 10^4.$
So ${\bf p}_1 = {\bf p}_2= 1/2; {\bf p}_{11}= {\bf p}_{12}= {\bf p}_{21}= {\bf p}_{22}= 1/2$ (symmetric state). 
Suppose we have fluctuations (\ref{*9}) with $\xi_{1i}^{(N)} \approx  \xi_{2i}^{(N)} \approx 1/2.$ 
Then $\epsilon_{1i}^{(N)}\approx \epsilon_{2i}^{(N)} \approx 500.$ 
So $n_{ij}=24\cdot 10^4 \pm 500.$ Hence, the relative deviation 
$\frac{\epsilon_{ji}^{(N)}}{m_{ji}}
=\frac{500}{25\cdot 10^4} \approx 0.002.$ 
Thus fluctuations of the relative magnitude $\approx 0,002$ 
produce the classical probabilistic rule.

It is evident that fluctuations of essentially larger magnitude 
\begin{equation}
\label{*9a} 
\epsilon_{1i}^{(N)}=2 \xi_{1i}^{(N)} (m_{1i})^{1/2} (m_{21})^{1/\alpha},\; 
\epsilon_{2i}^{(N)}= 2 \xi_{2i}^{(N)}(m_{2i})^{1/2}(m_{1i})^{1/\beta}, \alpha, \beta > 2, 
\end{equation}
where $\{ \xi_{1i}^{(N)}\}$ 
and $\{\xi_{2i}^{(N)}\}$ are bounded sequences $(N\rightarrow\infty),$ 
also produce (for ${\bf p}_{ij} \not= 0)$ the classical probability rule. 

{\bf Example 3.2.} Let all numbers $N, \ldots, m_{ij}$ be the same as in Example 3.1
and let deviations have behaviour (\ref{*9a}) with $\alpha=\beta=4.$ Here the relative deviation 
$\frac{\xi_{ij}^{(N)}}{mij} \approx 0,045.$

\section{Classical, quantum and `superquantum' physics}

In this section we find relations between different classes of physical experiments. 
First we consider so called classical and quantum experiments. Classical experiments 
produce the classical probabilistic rule (Bayes' formula). Therefore the corresponding
ensemble fluctuations have the asymptotic
$\delta_i^{(N)}\rightarrow 0, N\rightarrow\infty.$

Nevertheless, we cannot say that classical measurements give just a 
subclass of quantum measurements. In the classical domain we have no symmetric
relations ${\bf p}_{11}={\bf p}_{22}$ and ${\bf p}_{12}={\bf p}_{21}$. 
This is the special condition which connects the preparation procedures 
${\cal E}_1$ and ${\cal E}_2.$ This relation is a peculiarity of quantum 
preparation/measurement procedures.

Experiments with nonclassical probabilistic rules  are characterized by the condition
$\rm{\delta_i^{(N)}\not \rightarrow 0, N\rightarrow\infty.}$ Quantum experiments give
only a particular class of nonclassical experiments. Quantum experiments 
produce ensemble fluctuations of form (\ref{*7}), (\ref{*8}),
where coefficients $\xi_{1i}^{(N)}$ 
and $\xi_{20}^{(N)}$ satisfy (\ref{*8a}) and the orthogonality relation
\begin{equation}
\label{*h} 
\lim_{N\rightarrow\infty}
(\xi_{11}^{(N)}+\xi_{21}^{(N)})+\lim_{N\rightarrow\infty}(\xi_{12}^{(N)}+\xi_{22}^{(N)})=0\;.
\end{equation}

In particular, nonclassical domain contains (nonquantum) experiments
which satisfy condition of boundedness (\ref{*8a}), but not satisfy orthogonality relation (\ref{*h}). 
Here we have only the relation of quazi-orthogonality (\ref{*N2}). In this case the matrix of probabilities
is not double stochastic.
The corresponding probabilistic rule has the
form:
\begin{equation}
\label{*n7}
q_i={\bf p}_1{\bf p}_{1i}+{\bf p}_2{\bf p}_{2i}+2{\sqrt{{\bf p}_1{\bf p}_2{\bf p}_{1i}{\bf p}_{2i}}}\cos\theta_i.
\end{equation}
Here in general ${\bf p}_{11} + {\bf p}_{21}\not= 1, {\bf p}_{12} + {\bf p}_{22}\not= 1.$

We remark that, in fact, (\ref{*n7}) and (\ref{*N2}) imply that

$q_1={\bf p}_1 {\bf p}_{11}+{\bf p}_2 {\bf p}_{21} +2 \;\sqrt{{\bf p}_1 {\bf p}_2 {\bf p}_{11}{\bf p}_{21}}
\cos\theta_1;$

$q_2={\bf p}_1{\bf p}_{12}+{\bf p}_2{\bf p}_{22}- 2 \sqrt{{\bf p}_1{\bf p}_2{\bf p}_{11}{\bf p}_{21}}
\cos\theta_1.$

\section{Hyperbolic `quantum' formalism} 

Let us consider ensembles $\rm{S, T_1, T_2}$ such that ensemble
fluctuations have magnitudes (\ref{*7}), (\ref{*8}) where
\begin{equation}
\label{*8b}
|\xi_{1i}^{(N)}+\xi_{2i}^{(N)}|\geq 1+c, c>0,
N\rightarrow\infty.
\end{equation}
Here the coefficients $\lambda_i= \lim_{N\rightarrow\infty} (\xi_{1i}^{(N)}+\xi_{2i}^{(N)})$ can be
represented in the form $\lambda_i= \rm{ch} \;\theta_i, i=1,2.$ The corresponding probability rule 
is the following 

$q_i={\bf p}_1{\bf p}_{1i}+{\bf p}_2{\bf p}_{2i}+2
\sqrt{{\bf p}_1{\bf p}_2{\bf p}_{1i}{\bf p}_{2i}} \;\rm{ch} \; \theta_i, i=1,2.$

The normalization $q_1+q_2=1$ gives the orthogonality relation: 
\begin{equation}
\label{*N5} 
\sqrt{{\bf p}_{11}
{\bf p}_{21}}\; \rm{ch} \;\theta_1+\sqrt{{\bf p}_{12}{\bf p}_{22}} \;\rm{ch} \;\theta_2 = 0\;.
\end{equation}
Thus $\rm{ch} \theta_2=-\rm{ch} \theta_1 \sqrt{\frac{{\bf p}_{11} {\bf p}_{21}}{{\bf p}_{12} {\bf p}_{22}}}$ and,
hence, 

$q_2={\bf p}_1 {\bf p}_{12} + {\bf p}_2 {\bf p}_{22} - 2\; \sqrt{{\bf p}_1 {\bf p}_2{\bf p}_{11} {\bf p}_{21}}\;
\rm{ch}\theta_1.$ 

Such a
formalism can be called a {\it hyperbolic quantum formalism.} It describes a part of nonclassical reality
which is not described by `trigonometric quantum formalism'. Experiments (and preparation procedures
${\cal E}, {\cal E}_1, {\cal E}_2$) which produce hyperbolic quantum behaviour could be
simulated on computer. On the other hand, at the moment we have no `natural' physical phenomena
which are described by the hyperbolic quantum formalism. `Trigonometric quantum behaviour'
corresponds to essentially better control of properties in the process of preparation than
`hyperbolic quantum behaviour'. Of course, the aim of any experimenter is to approach
`trigonometric behaviour'. However, in principle there might exist such natural phenomena that
`trigonometric quantum behaviour' could not be achieved. In any case even the possibility of computer 
simulation demonstrates that quantum mechanics (trigonometric) is not complete 
(in the sense that not all physical reality is described by the standard quantum formalism). 
\footnote {We can compare the hyperbolic quantum formalism with the hyperbolic geometry.} 

{\bf Example 6.1.}
Let ${\bf p}_1=\alpha, {\bf p}_2=1-\alpha, {\bf p}_{11}=\ldots={\bf p}_{22}=1/2.$
Then 

$q_1= \frac{1}{2} +\sqrt{\alpha(1-\alpha)} \lambda_1, 
q_2={\frac{1}{2}}-\sqrt{\alpha(1-\alpha)}
\lambda_1.$ 

If $\alpha$ is sufficiently small, then $\lambda_1$ can be, in principle,
larger than 1:$\lambda_1= \rm{ch}\theta$.

\section{Quantum behaviour for macroscopic systems}

Our analysis shows that `quantum statistical behaviour' can be demonstrated by ensembles
consisting of macroscopic systems; for example, balls having colours $C=c_1,$ red, or $c_2,$ blue, and
weights $A=a_1=1$ or $a_2=2.$ Suppose that 
additional filters $F_i, i=1,2,$ produce fluctuations (\ref{*7}), (\ref{*8}), (\ref{*h}).
Then, instead of classical Bayes' formula (\ref{*1}),
we obtain  quantum probability rule (\ref{*2}). 

In the
context of the statistical simulation of quantum statistical behaviour via fluctuations (\ref{*7}), (\ref{*8})
(with (\ref{*h})) it would be useful to note that, in fact, we can choose constant coefficients 
$\rm{\xi_{ij}^{(N)}=\xi_{ij}.}$ Moreover, we have $\xi_{11}=-\xi_{12}$ and $\xi_{21}=-\xi_{22}.$ The latter is a
consequence of the general relations: 
\begin{equation}
\label{*R} 
\frac{\xi_{11}^{(N)}}{\xi_{12}^{(N)}}
\rightarrow-1,\; \;
\frac{\xi_{22}^{(N)}}{\xi_{21}^{(N)}} \rightarrow -1, N\rightarrow\infty.
\end{equation}
Asymptotic (\ref{*R}) can be obtained
from (\ref{*7}), (\ref{*8}):

{\bf Proof.} By (\ref{*7}) we have
\begin{equation}
\label{*q7} 
(n_{11}-m_{11})+(n_{12}-m_{12})=
2 \xi_{11} \sqrt{m_{11}m_{21}}
+2 \xi_{12}
\sqrt{m_{12}m_{22}}.
\end{equation}
The left hand side is equal to zero:
$(n_{11}+n_{12})-(m_{11}+ m_{12})=n_1^c
-n_1^c=0$
(as the ensemble $T_1$ has $n_1^c$ elements). Hence, by (\ref{*N1}) we get
$\xi_{11}=- \xi_{12}\; \sqrt{\frac{m_{12}}
{m_{21}} \frac{m_{22}}{m_{11}}} \rightarrow -\xi_{12}, N\rightarrow\infty\;$ 
(as ${\bf p}_{11}={\bf p}_{22}$ and ${\bf p}_{12}= {\bf p}_{21}).$
In the same way we obtain that 
$\xi_{21}= - \xi_{22} \; \sqrt{\frac{m_{12}}
{m_{21}} \frac{m_{22}}{m_{11}}} \rightarrow -\xi_{22}, N\rightarrow\infty\;.$

{\bf Conclusion.} {\it We demonstrated that so called quantum probabilistic rule has a
natural explanation in the framework of ensemble fluctuations induced 
by preparation procedures. In particular, the quantum rule for probabilities (with nontrivial
$\cos \theta$-factor) could be simulated for macroscopic physical systems via preparation
procedures producing the special ensemble fluctuations.}

\section{Appendix: correlations between preparation procedures}

In this section we study the frequency meaning of the fact that in the quantum formalism
the matrix of probabilities is double stochastic.
We remark that this
is a consequence of orthogonality of quantum states $\phi_1$ and $\phi_2$ 
corresponding to distinct values of a physical observable $C.$
We have
\begin{equation}
\label{*q8} 
\frac{{\bf p}_{11}}{{\bf p}_{12}}
=\frac{{\bf p}_{22}}{{\bf p}_{21}} \;.
\end{equation}

Suppose that (a), see section 2, is the origin of quantum behaviour.
Hence, all quantum features  are induced by the impossibility to create new ensembles $T_1$ 
and $T_2$ without to change properties of quantum particles. 
Suppose that,
for example, the preparation procedure 
${\cal E}_1$ practically destroys the property $A=a_1$
(transforms this property into the property $A=a_2)$.
So ${\bf p}_{11}=0.$ As a consequence, the ${\cal E}_1$ 
makes the property $A = a_2$ dominating.
So ${\bf p}_{12}\approx 1.$ 
Then the preparation procedure ${\cal E}_2$ {\it must} practically destroy
the property $A=a_2$ (transforms this property into the property $A=a_1)$.
So ${\bf p}_{22} \approx 0.$ 
As a consequence, the ${\cal E}_2$ makes 
the property $A=a_1$ dominating. So ${\bf p}_{21}\approx 1.$

Frequency relation (\ref{*N1}) can be represented in the following form:
\begin{equation}
\label{*M} 
\frac{m_{11}}{n_1^c} - \frac{m_{22}}{n_2^c} \approx 0,
N\rightarrow\infty\;.
\end{equation}
We recall that the number of elements in the ensemble $T_i$ is equal to $n_i^c.$

Thus 
\begin{equation}
\label{*M1} 
(\frac{n_{11} - m_{11}}{n_1^c}) -
(\frac{n_{22} - m_{22}}{n_2^c})
\approx \frac{n_{11}}
{n_1^c} - \frac{n_{22}}{n_2^c}.
\end{equation}
This is nothing than the relation between fluctuations of property
$A$ under the transition from the ensemble S to ensembles $T_1, T_2$ and distribution of this
property in the ensemble S.

\newpage

{\bf References}

[1] L. Accardi, The probabilistic roots of the quantum mechanical paradoxes.
{\it The wave--particle dualism. A tribute to Louis de Broglie on his 90th 
Birthday}, (Perugia, 1982). Edited by S. Diner, D. Fargue, G. Lochak and F. Selleri.
D. Reidel Publ. Company, Dordrecht, 297--330(1984).

[2] B. d'Espagnat, {\it Veiled Reality. An anlysis of present-day
quantum mechanical concepts.} (Addison-Wesley, 1995). 

[3] A. Peres, {\it Quantum Theory: Concepts and Methods.} (Kluwer Academic Publishers, 1994).

[4] J. S. Bell,
Rev. Mod. Phys., {\bf 38}, 447--452 (1966); J. F. Clauser , M.A. Horne, A. Shimony, R. A. Holt,
Phys. Rev. Letters, {\bf 49}, 1804-1806 (1969);
J. S. Bell, {\it Speakable and unspeakable in quantum mechanics.}
(Cambridge Univ. Press, 1987);
J.F. Clauser ,  A. Shimony,  Rep. Progr.Phys.,
{\bf 41} 1881-1901 (1978).

[5] L. Accardi, 
Phys. Rep., {\bf 77}, 169-192 (1981).
L. Accardi, A. Fedullo, 
Lettere al Nuovo Cimento {\bf 34} 161-172  (1982).
L. Accardi,
 Quantum theory and non--kolmogorovian probability.
In:   Stochastic processes in quantum theory and statistical physics,
ed. S. Albeverio et al., Springer LNP {\bf 173} 1-18 (1982).

[6] I. Pitowsky,  Phys. Rev. Lett, {\bf 48}, N.10, 1299-1302 (1982);
 S. P. Gudder,  J. Math Phys., {\bf 25}, 2397- 2401 (1984);
 S. P. Gudder, N. Zanghi,
 Nuovo Cimento B {\bf 79}, 291--301 (1984).

[7]  A. Fine,  Phys. Rev. Letters, {\bf 48}, 291--295 (1982);
P. Rastal, Found. Phys., {\bf 13}, 555 (1983).
W. De Baere,  Lett. Nuovo Cimento, {\bf 39}, 234-238 (1984);
{\bf 25}, 2397- 2401 (1984); 
 W. De Muynck, W. De Baere, H. Martens,
 Found. of Physics, {\bf 24}, 1589--1663 (1994);
W. De Muynck, J.T. Stekelenborg,  Annalen der Physik, {\bf 45},
N.7, 222-234 (1988).

[8] L. Accardi, M. Regoli,
Experimental violation of Bell's inequality by local classical variables.
To appear in: proceedings of the Towa Statphys conference, Fukuoka 8--11
November (1999), published by American Physical Society.

[9] L. Accardi, {\it Urne e Camaleoni: Dialogo sulla realta,
le leggi del caso e la teoria quantistica.} (Il Saggiatore, Rome, 1997).

[10] A.Yu. Khrennikov, {\it Non-Archimedean analysis: quantum
paradoxes, dynamical systems and biological models.}
(Kluwer Acad.Publ., Dordreht,  1997); {\it Interpretations of 
probability.} (VSP Int. Publ., Utrecht, 1999);
 J. Math. Phys., {\bf 41}, 1768-1777 (2000).

[11] J. Summhammer, Int. J. Theor. Physics, {\bf 33}, 171-178 (1994);
Found. Phys. Lett. {\bf 1}, 113 (1988); Phys.Lett., {\bf A136,} 183 (1989).

[12] R.  von Mises, {\it The mathematical theory of probability and
 statistics}. (Academic, London,  1964);
 
[13] A. N. Kolmogoroff, {\it Grundbegriffe der Wahrscheinlichkeitsrechnung.}
(Springer Verlag, Berlin, 1933); reprinted:
{\it Foundations of the Probability Theory}. 
(Chelsea Publ. Comp., New York, 1956).

[14]  L. E. Ballentine, {\it Quantum mechanics.} (Englewood Cliffs, 
New Jersey, 1989).

\end{document}